\documentclass[12pt,twocolumn]{article}
\usepackage{psfig}

\usepackage{fullpage}
\usepackage{psfig}

\newbox\grsign \setbox\grsign=\hbox{$>$}
\newdimen\grdimen \grdimen=\ht\grsign
\newbox\laxbox \newbox\gaxbox
\setbox\gaxbox=\hbox{\raise.5ex\hbox{$>$}\llap
        {\lower.5ex\hbox{$\sim$}}}\ht1=\grdimen\dp1=0pt
\setbox\laxbox=\hbox{\raise.5ex\hbox{$<$}\llap
        {\lower.5ex\hbox{$\sim$}}}\ht2=\grdimen\dp2=0pt
\def\simlt{\mathrel{\copy\laxbox}}

\begin{document}

\onecolumn
\begin{center}
{\large\bf Transients in the Local Universe}
\\\medskip
{S.~R.~Kulkarni$^{*}$ \& M.~M.~Kasliwal$^{*}$ \\
{\small $^{*}$California Institute of Technology, MS 105-24, Pasadena, CA 91125, USA}
}
\\\bigskip
{\it On behalf of the \\
 {\bf Palomar Transient Factory Collaboration} \& \\
 {\bf Large Synoptic Survey Telescope Transients Working Group}}
\\\bigskip
\end{center}
\bigskip
\noindent{\bf Submitted to the following Science Frontier Panels of Astro2010} 
\\
The Stars and Stellar Evolution \\
The Galactic Neighborhood \\
The Cosmology and Fundamental Physics \\
\\\bigskip
\noindent{\bf Disclosures} 
\\
This white paper was circulated to two groups:
the Palomar Transient Factory (PTF)\footnote{PTF is a project
whose only goal is a systematic investigation of variables and
transients in the optical sky. It is based around the Palomar
48-inch Oschin Schmidt telescope with a 96-million pixel detector and
7.8 square degrees field of view. PTF partners include California Institution of
Technology, Columbia University, Infrared Processing \&\ Analysis
Center, Lawrence Berkeley Laboratory, Las Cumbres Observatory Global
Telescope, Weizmann Institute of Science \&\ UC Berkeley. PTF achieved
first light in December 2008 and expects to be in routine operation
by May 2009.} and the Science Working Group on Transients of the Large 
Synoptic Survey Telescope (LSST)\footnote{The LSST project is based around a 8.4-m
telescope coupled to a 3-billion pixel camera. LSST plans
to survey the entire sky visible from Chile in multiple colors
every week. The goals range from near-Earth asteroids to
cosmology. Twenty eight institutions are members of this
consortium. LSST hopes to see first light in 2014 and start transient
search in 2015.}.
The two authors have incorporated most of the feedback.
However, the final version is the responsibility of the
two authors and is not legally binding on the PTF and LSST collaborations.

\bigskip
\bigskip
\noindent We thank E.~S.~Phinney, L.~Wen, M.~Shara, C.~Fryer, D.~Fox, T.~Tyson, J.~Bloom, 
A.~Gal-Yam, C.~Ott, H.~Bond, A.~Rau, E.~O.~Ofek and R.~Quimby for feedback and 
support of this science.
\\\bigskip

\noindent The first author is the Principal Investigator of PTF project and
the Chair of the Science Working Group on Transients of LSST. The second 
author is a graduate student whose thesis is centered
around the topic of this white paper.

\clearpage

\begin{twocolumn}

\begin{abstract}
Two different reasons make the search for transients in the nearby
Universe (d$\simlt200\,$Mpc) interesting and urgent. First, there
exists a large gap in the luminosity of  the brightest novae
($-10\,$mag) and  that of sub-luminous supernovae ($-16$\,mag).
However, theory and reasonable speculation point to several potential
classes of objects in this ``gap''.  Such objects are best found
in the Local Universe.  Second, the nascent field of Gravitational
Wave (GW) astronomy and the budding fields of Ultra-high energy
cosmic rays, TeV photons, astrophysical neutrinos
 are likewise limited to the Local Universe by physical effects
(GZK effect, photon pair production) or instrumental sensitivity
(neutrino and GW). Unfortunately, the localization of these
new telescopes is poor and precludes identification of the
host galaxy (with attendant loss of distance and physical
diagnostics). Both goals can be met with wide field imaging
telescopes acting in concert with follow-up telescopes. Astronomers
must also embark upon completing the census of galaxies in 
the nearby Universe.

\end{abstract}

\section{A Historical Summary}

Variable objects have played a major role in the history
of astronomy.  At the beginning of the previous century, the pulsating
Cepheids showed that the Galaxy was much larger than had been assumed
before.
Hubble's discovery of Cepheids in the Andromeda nebula showed that it
was a galaxy similar in size to our own.

F. Zwicky's 18-inch Palomar Schmidt program was the first systematic
study of the transient sky.  Zwicky
laid the foundation of the field of supernovae which in turn
touched upon the origin of cosmic rays, the synthesis of Iron and
higher-$(A,Z)$ elements and the collapse of stellar cores. This success
led to the 48-inch Palomar Oschin Schmidt telescope,
famous for two Sky Surveys.

At the end of the previous century, supernovae came back into main
stream. The first indication of a new constituent of the Universe,
dark energy, was deduced from the dimming of Ia supernovae located
at cosmological distances (Perlmutter et al. 1999, Riess et al.
1998).
The last decade saw a veritable explosion 
in the field of gamma ray bursts -- the most relativistic explosions
in Nature. 

The purpose of this white paper is to explicitly draw attention to
the importance of a systematic study of transients in the local
Universe (defined here as distance $\simlt 200\,$Mpc). The 
rationale for this myopic view is laid in \S\ref{sec:TheGap} and
\S\ref{sec:NewAstronomy}. We argue that the path to discovery
necessarily lies through identification and elimination of
fore- and back-ground transient events (\S\ref{sec:PilotPrograms}).
We discuss upcoming facilities in (\S\ref{sec:PTF}).
While detection of any cosmic source by LIGO constitutes a major
step there is wide-spread agreement of the necessity of
arc-second localization (which means in essence detecting an electromagnetic
counterpart). We end with a proposed plan to identify electromagnetic
counterparts to LIGO sources (\S\ref{sec:LIGO}).

Here, we limit the discussion to transients in the local Universe. We
refer the reader to the following white papers: Cosmological explosions
(Wozniak et al.), gravitational wave sources (Bloom et al., Phinney et al.)
and Galactic variable stars (Becker et al.).

\section{Rationale \&\ Motivation: Events in the Gap}
\label{sec:TheGap}

\begin{figure}[hbt]
\centerline{
\psfig{figure=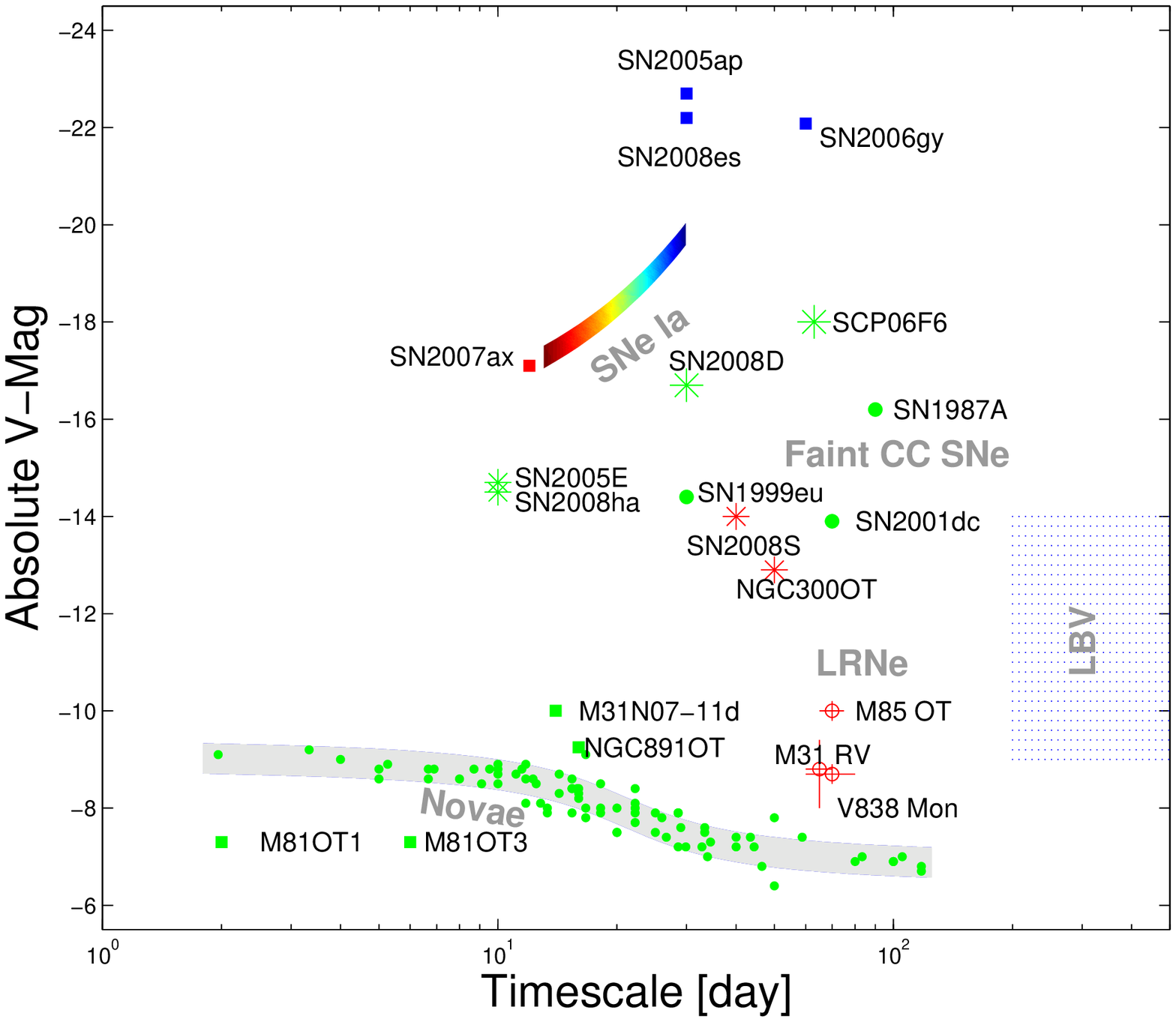,width=8.5cm}}
\caption[]{
\scriptsize
The phase space of cosmic transients : peak $V$-band luminosity as a function
of duration, with color a measure of the true color at maximum.
Shown are the known explosive (supernovae) and eruptive (novae,
luminous blue variables [LBV]) transients.  Also shown are new types
of transients (all found over the last two years): 
the peculiar transients M\,85\,OT2006-1, M31-RV and V838\,Mon, which
possibly form a new class of ``luminous red novae''; almost every
scenario has been suggested -- core collapse, common envelope event,
planet plunging into star, a peculiar nova and a peculiar AGB phase;
the baffling
transient with a spectrum of a red-shifted Carbon star, SCP\,06F6
(Barbary et al 2008; see also Soker et al 2008);
a possible accretion induced collapse event SN\,2005E (Perets et al. 2009);
the extremely faint, possibly Type Ibc, SN\,2008ha (Valenti et al. 2008);
and peculiar eruptive events with extremely red progenitors
SN\,2008S and NGC300-OT (Thompson et al. 2008, Smith et al. 2008, Bond et al. 2009). [Figure adapted from Kulkarni et al. (2007).]

}
\label{fig:taumv}
\end{figure}

A plot of the peak luminosity versus a duration that is characteristic
(based on physics or convention) is a convenient way to summarize
explosive events.  We first focus on novae and supernovae of 
type Ia (SN\,Ia). As can be seen from Figure~\ref{fig:taumv} novae and
SN\,Ia form a distinctly different locus. Brighter
supernovae take a longer time to evolve (the ``Phillips'' relation;
Phillips 1993) whereas  the opposite is true of novae:
the faster the nova decays the higher the luminosity (the ``Maximum
Magnitude Rate of Decline'', MMRD relation; see, for example, Della
Valle and Livio 1995, Downes and Duerbeck 2000).

The primary physical parameter in SN\,Ia is the amount of
Nickel that was synthesized. There is almost a factor of 10 variation
between the brightest (``1991T-like'') and the dimmest
(``1991bg-like'') SN\,Ia. The Phillips relation has been quantified with
high precision and the theory is well understood. In contrast, the
MMRD does not enjoy the same quantity or quality of light curves
as those of Ia supernovae. Fortunately, dedicated on-going nova searches in
M31 and the P60-FasTING project (see \S\ref{sec:PilotPrograms}) 
has vastly increased the number of well-sampled light curves.

The large gap between the brightest novae (say $M_{\rm V} \sim -10$)
and the faintest supernovae ($M_{\rm V}\sim -16$), especially
on timescale shorter than thirty days, is hard to miss.
The gap beckons astronomers to search the heavens for new populations.

A discussion of potential new classes of events in the gap would
benefit from a review of the basic physics of explosions. 
An important
complication is a potential heat source at the center: a hot white
dwarf (novae) or gradual release of radioactive energy (supernovae).

The primary physical parameters are:  the mass of the ejecta ($M_{\rm
ej}$), the velocity of the ejecta ($v_s$), the radius of the
progenitor star ($R_0$) and the total energy of the explosion
(${\mathcal{E}}_0$).  Two distinct sources of energy contribute to
the explosive energy: the kinetic energy of the ejecta, ${\mathcal{E}_{\rm
k}}\equiv (1/2) M_{\rm ej} v_{s}^2$ and the energy in the photons
(at the time of the explosion), $\mathcal{E}_{\rm ph}$.

Assuming spherical symmetry and homogeneous density, the
following equation describes the gains and losses suffered
by the store of heat ($E$):
\begin{equation} 
\dot E = \varepsilon(t)M_{\mathrm{ej}} - L(t) - 4\pi R(t)^2 P v(t).
\label{eq:dotE}
\end{equation}
Here, $L(t)$ is the luminosity radiated at the surface and
$\varepsilon(t)$ is heating rate (energy per unit time) per gram from any source of energy
(e.g.\ radioactivity or a long-lived central source).  $P$ is the
total pressure and is given by the sum of gas and photon pressure.

Next,  we resort to the so-called ``diffusion'' approximation (see
Arnett 1996; Padmanabhan 2000, volume II, \S 4.8),
\begin{equation}
L = E_{\rm ph}/t_{\rm d},
\label{eq:L}
\end{equation}
where  $E_{\rm ph}=aT^4V$ is the energy in
photons  [$V$ is the volume, $(4\pi/3)R^3$], and
\begin{equation}
t_d=B\kappa M_{\rm ej}/cR
\label{eq:td}
\end{equation} 
is the timescale for a photon to diffuse from the center
to the surface. The pre-factor $B$ in Equation~\ref{eq:td} depends
on the geometry and, following Padmanabhan, we set
$B=0.07$.  $\kappa$ is the mass opacity.

We will make one simplifying assumption: most of the acceleration
of the ejecta takes place on the initial hydrodynamic timescale, $\tau_h=R_0/v_s$,
and subsequently coasts at $R(t) = R_0 + v_{\rm s} t$.

First, let us consider a ``pure'' explosion i.e. no subsequent
heating [$\varepsilon(t)=0$].  If photon pressure dominates then
$P=1/3 (E/V)$  and an analytical formula for $L(t)$ can be obtained
(Padmanabhan, {\it op cit}):
\begin{equation}
L(t) = L_0 \exp\bigg(-\frac{t\tau_h + t^2/2}{\tau_h\tau_d}\bigg);
\label{eq:Ltphot}
\end{equation}
here,  $\tau_d=
B(\kappa M_{\rm ej}/cR_0)$ is the initial diffusion timescale and
$L_0={\mathcal{E}}_{\rm ph}/\tau_d$.  

From Equation~\ref{eq:Ltphot} one can see that the 
light curve is divided into (1) a plateau phase which
lasts until about $\tau=\sqrt{\tau_d\tau_h}$ after which (2) the
luminosity undergoes a (faster than) exponential decay.  The duration
of the plateau phase is
	\begin{equation} 
	\tau = \sqrt{\frac{B\kappa M_{\rm ej}}{cv_s}}
	\label{eq:tau} 
	\end{equation}
and is independent of $R_0$. The plateau luminosity is
	\begin{eqnarray}
		L_p &=& {\mathcal{E}}_{\rm ph}/\tau_d =
		\frac{cv_s^2 R_0}{2B\kappa}\frac{\mathcal{E}_{\rm
		ph}}{\mathcal{E}_{\rm k}}.  \label{eq:Lp}
	\end{eqnarray}
As can be seen from Equation~\ref{eq:Lp} the peak luminosity is
independent of the mass of the ejecta but directly proportional to
$R_0$. To the extent that there is rough equipartition\footnote{This
is a critical assumption and must be checked for every potential
scenario under consideration. In a relativistic fireball most of the energy
is transferred to matter. For novae, this assumption is violated (Shara, pers. comm.).} between the kinetic energy and the
energy in photons, the luminosity is proportional to the square of
the final coasting speed, $v_s^2$.

Pure explosions satisfactorily account for supernovae of type IIp.
Note that since $L_p \propto R_0$ the larger the star the higher
the peak luminosity.  SN~2006gy, one of the brightest supernovae,
can be explained by invoking an explosion in a ``star'' which is
much larger (160\,AU) than any star (likely the material shed by a
massive star prior to its death; see Smith \&\ McCray 2007).

{\it Conversely}, pure explosions resulting from the deaths of compact
stars (e.g.  neutron stars, white dwarfs or even stars with radius
similar to that of the Sun) will be very faint.
For
such progenitors visibility in the sky would require some sort of
additional subsequent heat input and discussed next.

First we will consider ``supernova''-like events i.e. events in
which the resulting debris is heated by radioactivity. One can
easily imagine a continuation of the Ia supernova sequence.  We
consider three possible examples for which we expect a smaller
amount of radioactive yield and a rapid decay (timescales of days):
coalescence of compact objects, accreting white dwarfs (O-Ne-Mg) and
final He shell flash in AM CVn systems.

Following Li \&\ Paczynksi (1998), Kulkarni (2005) considers the
possibility of the debris of neutron star coalescence being heated
by decaying neutrons. Amazingly (despite the 10-min decay time of
free neutrons)  such events (dubbed as ``macronovae'') are detectable
in the nearby Universe over a period as long as a day, provided
even a small amount ($\sim >10^{-3}\,M_\odot$) of free neutrons
is released in such explosions. 

Bildsten et al. (2007) consider a Helium nova (which arise in AM
CVn systems). For these events (dubbed ``.Ia'' supernovae), not
only radioactive Nickel but also radioactive Iron is expected.

Intermediate mass stars present two possible paths to sub-luminous
supernovae. The O-Ne-Mg cores could either lead to a disruption (bright
SN but no remnant) or a sub-luminous explosion (Kitaura et al. 2006).
Separately, the issue of O-Ne-Mg white dwarfs accreting matter from a companion
continues to fascinate astronomers. 
The likely possibility is a neutron star but the outcome
depends severely on the unknown effects of rotation and magnetic
fields. One possibility is an explosion with low Nickel yield (see Metzger
et al. 2008 for a recent discussion and review of the literature).

An entirely different class of explosive events are expected to
arise in massive or large stars: birth of black holes (which can
range from very silent events to GRBs and everything in between),
strong shocks in super-giants (van den Heuvel 2008) and common
envelope mergers.  Equations~\ref{eq:tau} and \ref{eq:Lp} provide
a guidance to the expected appearance of such objects. Fryer et al.
2007 develop detailed model for faint, fast supernovae due to Nickel
``fallback'' into the black  hole. For the case of the birth of a black 
hole with no resulting radioactive
yield (the newly synthesized material could be advected into the
black hole) the star will slowly fade away on a timescale of
min($\tau_d,\tau$). Modern surveys are capable of finding such wimpy
events (Kochanek et al. 2008).

In the spirit of this open-ended discussion
of new transients we also consider the case where the gas pressure
could dominate over photon pressure. This is the regime of weak
explosions.  If so, $P=2/3 (E/V)$ and Equation~\ref{eq:dotE} can
be integrated to yield 
	\begin{equation} L(t) = \frac{L_0}{(t/\tau_h+1)}
	\exp\bigg(-\frac{\tau_h t + t^2/2}{\tau_h \tau_d}\bigg).  
	\label{eq:Ltgas}
	\end{equation} 
In this case the relevant timescale is the hydrodynamic timescale.
This regime is populated by Luminous Blue Variables and Hypergiants.
Some of these stars are barely bound and suffer from bouts of
unstable mass loss and photometric instabilities.

As can be gathered from Figure~\ref{fig:taumv}
the pace of discoveries over the past two years gives
great confidence to our expectation of filling in the phase
space of explosions.

\section{Rationale \&\ Motivation: New Astronomy}
\label{sec:NewAstronomy}

Four entirely different fields of astronomy are now being nurtured
by physicists. For these
fields all that matters is transients in the local Universe.

Cosmic rays with energies exceeding $10^{20}\,$eV are strongly
attenuated owing to the production of pions through interaction
with the cosmic micro-wave background (CMB) photons [the famous
Greisen-Zatsepin-Kuzmin (GZK) effect]. Recently, the Pierre Auger
Observatory (PAO 2007) has found evidence showing that such cosmic
rays with energies above $6\times 10^{19}\,$eV are correlated with
distribution of matter in the local 75-Mpc sphere.  

Similarly very high energy (VHE) photons (TeV and PeV) have a highly
restricted horizon. The TeV photons interact with CMB photons and
produces electron-positron pairs. A number of facilities are now
routinely detecting extra-galactic TeV photons from objects in the
nearby Universe (Veritas, MAGIC, HESS, CANGAROO).

Neutrino astronomy is another budding field with an expected vast
increase in sensitivity. The horizon here is primarily limited by
sensitivity of the telescopes (ICECUBE).  

Gravitational wave (GW) astronomy suffers from both poor localization
(small interferometer baselines) and sensitivity.
The horizon radius
is 50 Mpc for  \texttt{enhanced} LIGO (e-LIGO) and about 200\,Mpc 
for \texttt{advanced}
LIGO (a-LIGO) to observe neutron star coalescence.
There is wide spread agreement
that the greatest gains will require electromagnetic localization
of the event.  

Progress in these areas, especially GW astronomy,  {\it require} arc-second localization.
Unfortunately, all sorts of foreground and background
transients {\it will} be found within the several to tens of sq deg
of expected localizations.  Studying
each of these transients will result in significant ``opportunity
cost''.  This issue is addressed in \S\ref{sec:LIGO}.

\section{Foreground Fog \&\ Background Haze}
\label{sec:PilotPrograms}

Ongoing projects of modest scope offer a glimpse of 
the pitfalls on the road to 
achieving the grand goals summarized in 
\S\ref{sec:TheGap} and
\S\ref{sec:NewAstronomy}. Nightly monitoring of M31 for novae (several groups)
and our Palomar 60-inch program of nearby galaxies already show high
variance in the MMRD relation (Figure~\ref{fig:novae}).
The large scatter of the new novae suggests
that in addition to the mass of the white dwarf, other physical
parameters play a role (such as accretion rate, white dwarf luminosity, e.g. Shara 1981).

A nightly targeted search of nearby rich clusters (Virgo, Coma
and Fornax) using the CFHT (dubbed ``COVET'') 
and the 100-inch du Pont (Rau et al. 2008a) telescopes  
uncover the extensive fore-ground fog 
(asteroids, M dwarf flares, dwarf novae) and the background
haze (distant, unrelated SN). The pie-chart in Figure~\ref{fig:covetpie}
dramatically illustrates that {\it new discoveries require efficient
elimination of fore- and back-ground events.}

\begin{figure}[hbt]
\centerline{
\psfig{figure=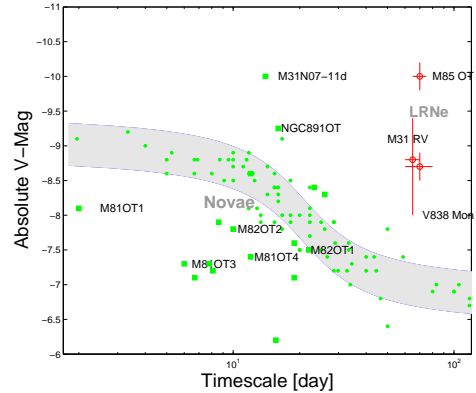,width=7cm}}
\caption[]{
\scriptsize
A plot of the peak absolute magnitudes versus decay timescale of
novae discovered by the Palomar P60-FasTING project. The shaded grey region represents
the Maximum Magnitude Rate of Decline relationship bounded by $\pm
3\sigma$ (Della Valle and Livio 1995). The data that defined this
MMRD are shown by green circles. Squares indicate novae discovered 
by P60-FasTING in 2007-2008 (unlabelled if in M31). 
[Preliminary results from Kasliwal et al. 2009b, in prep.]
}
\label{fig:novae}
\end{figure}

\begin{figure}[hbt]
\centerline{
\psfig{figure=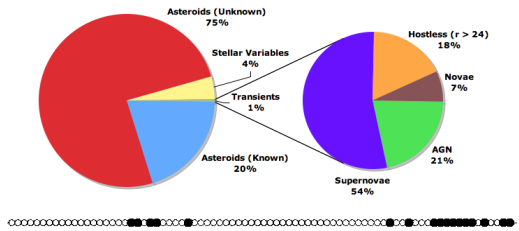,width=7cm}}
\caption[]{
\scriptsize
28 COVET transients were discovered during our
pilot run in 2008A (7 hours) -- two novae and
the remainder background supernovae and AGN. Transients with no point source or galaxy
host to a limiting magnitude of r$>$24 are classified as hostless.
Of the 2800 candidates, our pipeline automatically
rejected 99\% as solar-system or galactic. 
[Preliminary version from Kasliwal et al.  2009c, in prep.]}
\label{fig:covetpie} 
\end{figure}

\section{The Era of Synoptic Imaging Facilities}
\label{sec:PTF}

There is widespread agreement that we are now on the
threshold of the era of synoptic and
wide field imaging at optical wavelengths. This is
best illustrated by the profusion of operational
(Palomar Transient Factory, PanSTARRS), imminent
(SkyMapper, VST, ODI) and future facilities (LSST)
in optical and SASIR in infrared.

In Table~\ref{tab:rates} and Figure~\ref{fig:rates} we present
current best estimates for the rates of various events and the
``grasp'' of different surveys.

\begin{figure}[!hbt]
\centerline{
\psfig{figure=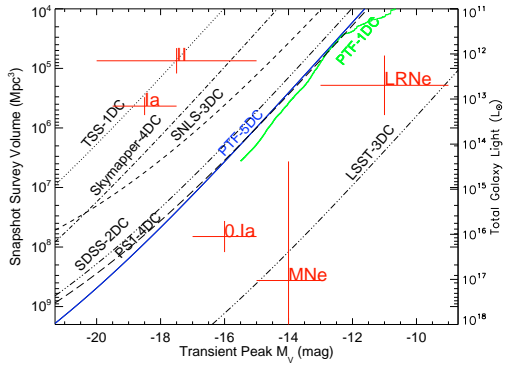,width=7cm,angle=90}}
\caption[]{
\scriptsize
Volume probed by various surveys (in a single specified cadence period) 
as a function of transient absolute
magnitude. Red plusses represent the minimum survey volume needed
to detect a single transient event (the uncertainty in the y-axis
is due to uncertainty in rates).  PTF-5d (blue-solid) is more
sensitive than TSS (dotted), Skymapper (dot-dashed), Supernova
Legacy Survey (SNLS, dashed) and SDSS-SN (double-dot dashed) and
competitive with PanSTARRS-1 (PS1-MD, long dashed).  Lines for each
survey represent one transient event in specified cadence period.
PTF-1d (green solid line) represents a targeted 800 sq deg survey 
probing luminosity concentrations in
the local Universe, with a factor of three larger effective survey
volume than a blind survey with same solid angle. PTF will discover
hundreds of supernovae and possibly several rare events such as
``0.Ia'',  Luminous Red Novae (LRNe) and Macronovae (MNe) per year. The 
LSST (Wide Fast Survey) will discover hundreds of these rare events. 
[Adapted from figure by Bildsten et al. 2009.]

}
\label{fig:rates} 
\end{figure}

\begin{table*}[ht]
\begin{center}
\caption{Properties and Rates for Optical Transients$^a$}
\begin{footnotesize}
\begin{tabular}{lcccccl}
\hline
\hline
Class           &  M$_v$ & $\tau$$^b$ & Universal Rate (UR) & PTF Rate & LSST Rate \\
& [mag] & [days] &  & [yr$^{-1}$] &  [yr$^{-1}$] \\
\hline
Luminous red novae & $-9..-13$ & 20..60 & $(1..10)\times10^{-13}$\,yr$^{-1}$ L$_{\odot,K}^{-1}$ & 0.5..8 & 80..3400 \\
Fallback SNe    & $-4..-21$ & $0.5..2$ &  $<5\times10^{-6}$\,Mpc$^{-3}$ yr$^{-1}$  &  $<$3 & $<$800  \\
Macronovae      & $-13..-15$ & 0.3..3 & $10^{-4..-8}$\,Mpc$^{-3}$ yr$^{-1}$ &  0.3..3 & 120..1200 \\
SNe~.Ia       & $-15..-17$ & 2..5 & $(0.6..2)\times10^{-6}$\,Mpc$^{-3}$ yr$^{-1}$ &  4..25 & 1400..8000 \\
SNe~Ia          & $-17..-19.5$ & 30..70 & $^c$ $3\times10^{-5}$\,Mpc$^{-3}$ yr$^{-1}$ & 700 & 200000$^d$ \\
SNe~II & $-15..-20$ & 20..300 & $(3..8)\times10^{-5}$\,Mpc$^{-3}$ yr$^{-1}$  & 300 & 100000$^d$ \\
\hline
\end{tabular}
\begin{itemize}
$^a${Table from Rau et al. 2008b; see references therein.} 
$^b${Time to decay by 2 magnitudes from peak.} 
$^c${Universal rate at z$<0.12$.} 
$^d${From M. Wood-Vassey, pers. comm.}
\end{itemize}
\label{tab:rates}
\end{footnotesize}
\end{center}
\end{table*}

The reader should be cautioned that many of these rates are very rough. 
Indeed, the principal goal of the Palomar Transient
Factory is to accurately establish the rates of the fore- and
back-ground events.
It is clear from Figure~\ref{fig:rates} that {\it the impressive
grasp of LSST is essential to the discovery of rare events.}

\section{Localizing LIGO events}
\label{sec:LIGO}

\begin{table*}[!hbt]
\begin{center}
\begin{footnotesize}
\caption{Galaxy Characteristics in LIGO Localizations \label{tab:galchar}}
\begin{tabular}{lcccccc}

\hline
 & & E-LIGO  &  & &  A-LIGO  & \\
\hline
 & 10\% & 50\% & 90\% & 10\% & 50\% & 90\% \\ 
\hline
GW Localization (sq deg) & 3 & 41 & 713 & 0.2 & 12  & 319 \\
Galaxy Area (sq arcmin) & 4.4 & 26 & 487 & 0.15 & 20.1 & 185 \\
Galaxy Number & 1 & 31 & 231 & 1 & 76 & 676 \\
Galaxy Luminosity (Log) & 10.3 & 11.2 & 12.1 & 10.9 & 12.0 & 13.0  \\
\hline
%
\end{tabular}
\end{footnotesize}
\end{center}
\end{table*}

\begin{figure}[!hbt]
\centerline{
\psfig{figure=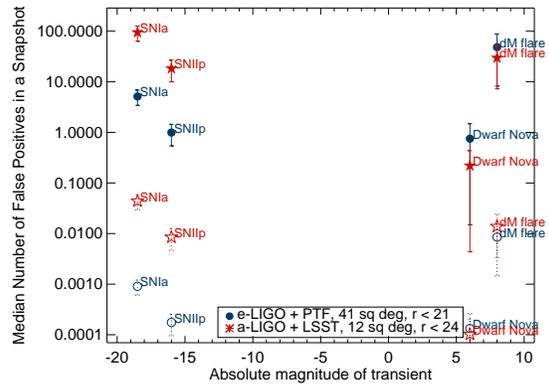,angle=90,width=8cm}}
\caption{\scriptsize Number of false positives in a single snapshot. 
For e-LIGO (blue circles), we use the median localization of
41 sq deg and follow-up depth of  r$<$21 is assumed. For a-LIGO
(red stars), we use the median localization of 12 sq deg and follow-up depth 
of r$<$24. Filled symbols denote false positives in the entire
error circle and empty symbols show false positives that are spatially
coincident with nearby galaxies. Dwarf novae and 
M-dwarf (dM) flares constitute the foreground fog and the ``error bars'' on 
numbers represents the  dependence on galactic latitude. Supernovae (Ia,IIp) 
constitute background haze.}
\label{fig:false-positive}
\end{figure}

We simulated a hundred GW events (Kasliwal et al. 2009a, in prep) and compute the 
exact localization on sky (assuming a neutron-star neutron-star merger 
waveform and triple coincidence data from LIGO-Hanford, LIGO-Louisiana and Virgo). 
The localizations range between 3--700 sq deg for e-LIGO and 0.2--300 sq deg 
for a-LIGO (range quoted between 10th and 90th percentile). The Universe is 
very dynamic and the number of false positives in a single snapshot
is several tens for a median localization  (see Figure~\ref{fig:false-positive}).

Fortunately, the sensitivity-limited $<200\,$Mpc horizon of GW astronomy
is a blessing in disguise. The opportunity cost (\S\ref{sec:PilotPrograms})
can be substantially reduced by
restricting follow-up to those transients that are spatially coincident with 
galaxies within $200\,$Mpc. Limiting the search to the area covered by galaxies 
within a LIGO localization reduces a sq deg problem to a sq arcmin problem 
 --- a reduction in false positives by three orders of magnitude! 

Given the total galaxy light in the localization, 
we also find that the number of false positives due to unrelated supernovae or novae within 
the galaxy is negligible. 
To be sensitive to transients as faint at peak as M\,$<$\,$-$13 (fainter than the faintest observed
short hard gamma ray burst optical afterglow), e-LIGO needs at-least a 1-m class telescope for 
follow-up (m\,$<$\,21, 50\,Mpc) and a-LIGO an 8-m class (m\,$<$\,24, 200\,Mpc). 
Given the large numbers of galaxies within the localization (Table~\ref{tab:galchar}),
a large field of view camera ($>$\,5 sq deg) will help maximize depth and cadence as compared
to individual pointings. Thus, PTF is well-positioned
to follow-up e-LIGO events and LSST to follow-up a-LIGO events.  

The feasibility hinges on a complete galaxy catalog. We compiled all available distances 
to nearby
galaxies and find that this catalog is only 70\% complete at e-LIGO distance 
and 55\% complete at a-LIGO distance.
It is important that astronomers
embark upon completing the census of galaxies within 200\,Mpc.
\\\bigskip
\\\smallskip
\noindent {\bf References} 
\\\scriptsize
\noindent Arnett D. 1996, Supernova and Nucleosynthesis (Princeton University Press) \\
Barbary K. et al. 2008, ApJ, accepted, arXiv:0809:1648 \\
Bildsten L. et al. 2007, ApJ, 662, 95 \\
Bond, H. E. el al. 2009, astroph, arXiv:090.0198 \\
Della Valle M. \& Livio 1995, ApJ, 452, 704 \\
Downes R.A. \& Duerbeck 2000, AJ, 120, 2007 \\
Fryer C. et al. 2007, ApJ, 662, 55 \\
Kasliwal, M.M. et al. 2008, ApJ, 638L, 29 \\
Kitaura, F. S. et al. 2006, A\&A 450, 345\\
Kochanek, C. et al. 2008, ApJ, 684, 1336 \\
Kulkarni S.R. 2005, astro-ph/0510256 \\
Kulkarni, S. R. \&\ Rau, A. 2006, ApJ, 644, L63 \\
Kulkarni S.R. et al. 2007, Nature, 447, 458 \\
Li L. \&\ Paczynski, B. 1998, ApJ, 507, 59 \\
Matsuoka M. et al. 2007, SPIE, 6686, 32 \\
Metzger B. et al. 2008, MNRAS, submitted, arXiv:0812:3656 \\
PAO, Pierre Auger Observatory, http://www.auger.org \\
Padmanabhan P. 2000, Theoretical Astrophysics (Cambridge University Press) \\
Perlmutter S. et al. 1999, ApJ, 517, 565 \\
Phillips M.M. 1993, ApJ, 413, L105 \\
Rau A. et al. 2008b, PASP, submitted \\
Rau A. et al. 2008, ApJ, 682, 1205 \\
Riess, A.G. et al. 1998, AJ, 116, 2009 \\
Shara, M. 1981, ApJ, 243, 926 \\
Smith N. et al. 2007, ApJ, 671, L17 \\
Smith N. et al. 2008, ApJL, submitted, arXiv:0811:3929 \\
Soker N. et al. 2008, submitted, arXiv:0812:1402 \\
Thomspon T.A. et al. 2008, ApJ, submitted, arXiv:0809:0510 \\
van den Heuvel E.P.J. 2008, Nature, 451, 775 \\
Valenti, S. et al. 2009, submitted, Nature, arXiv:0901:2074 \\

\end{twocolumn}
\end{document}